\documentclass{article} 
\usepackage{times}  
\usepackage{helvet} 
\usepackage{courier}  
\usepackage[hyphens]{url}  
\usepackage{graphicx} 
\usepackage{graphicx}  
\usepackage{amssymb}
\usepackage{amsmath}
\usepackage[ruled,vlined]{algorithm2e}
\usepackage[sl]{caption}
\usepackage{tikz}
\usepackage{tikzscale}
\usepackage{pgfplots}
\usepackage{subfigure}

\title{LAC-Nav: Collision-Free Mutiagent Navigation Based on The Local Action Cells}

\author{Li Ning and Yong Zhang\\ Shenzhen Institutes of Advanced Technology, Chinese Academy of Science\\\tt{\{li.ning, zhangyong\}@siat.ac.cn}}

\begin{document}
\maketitle

\begin{abstract}
  Collision avoidance is one of the most primary problems in the
  decentralized multiagent navigation:
  while the agents are moving towards their own targets,
  attentions should be paid to avoid the collisions with the others.
  In this paper, we introduced the concept of the \emph{local action cell},
  which provides for each agent a set of velocities that are safe to perform.
  Consequently, as long as the local action cells are updated on time and
  each agent selects its motion within the corresponding cell,
  there should be no collision caused.
  Furthermore, we coupled the local action cell with an adaptive learning
  framework, in which the performance of selected motions are evaluated and used
  as the references for making decisions in the following updates.
  The efficiency of the proposed approaches were demonstrated
  through the experiments for three commonly considered scenarios,
  where the comparisons have been made with several well studied strategies.
\end{abstract}

\section{Introduction}

Collision-free navigation is a fundamental and important problem
in the design of the multiagent systems, which are widely applied
in the fields such as robots control and traffic engineering.
When moving the agents in an environment with static or dynamic obstacles,
it is usually a necessary requirement to
well plan the trajectories such that no collision is caused.
As the number of agents increases and the environment area becomes large,
planning the realtime motions for all agents in the centralized manner
causes huge amount of the calculations, and is often restricted
by the efficiency of the communication between the agent and the planning monitor.
Therefore, it is natural (sometimes necessary) to consider
the decentralized navigation approaches, by which the individual agent
is responsible for sensing the nearby obstacles and performing the
proper motion to progress towards its destination without causing any collisions.
On the other hand, as a consequence of the decentralized navigation,
it is in general difficult for the agents to fully coordinate
before making the independent moves.
Thus they should also be considered and avoided as the obstacles to each other.

As noticed in the existing works, when avoiding collisions with the other agents,
it is important to take into account the fact that they are also intelligent
to perform the collision avoidance motions
(otherwise, undesirable oscillations may be observed during the navigation).
Consequently, it is not necessary for any individual agent
to take all the responsibility of making sure that the performed motion is safe.
ORCA \cite{Berg2011} is a well known decentralized approach
that guarantees to generate the optimal reciprocal collision-free velocities,
except for some certain conditions with densely packed agents.
BVC \cite{Zhou2017} has been proposed to restrict the agents moving inside
the non-intersecting areas, and thus the collision avoidance is guaranteed.
After the safe field of the motions
(i.e.\ the safe range of velocities or the safe area of positions) is determined,
both of the ORCA-based approaches and the BVC-based approaches
usually select the motion that is closest to the preferred motion,
within the safe field. Such a greedy strategy is natural and
widely used in the local-search-based optimizations.
However, it may cause the less efficient performance in the multiagent
navigation, as the agents may refuse to detour until there is no chance
to approach the target. In the worst case, with the greedy selection,
agents may get stuck in a loop of two or more situations (also known as the deadlocks).
Although some tricks have been proposed to fix such drawbacks
(including the ideas described in \cite{Zhou2017}),
they are not always valid in the concrete implementations,
and the improvements vary from case to case.

In this work, in order to improve the navigation efficiency,
we extend the buffered Voronoi cell \cite{Zhou2017} in the velocity space,
and consider the relative velocities for their effects on causing the potential conflicts.
In the selection of the motion to perform, the traveling progress has been also considered,
and consequently the agents may detour earlier, as long as approaching directly to the target
leads to less progress in the moving distance.

\paragraph{Problem formation.}
In this work, we consider a set $\mathcal{A}$ of the disk-shaped agents moving in the plane.
For any time point, agent $a_i \in \mathcal{A}$ of position $p_i \in \mathbb{R}^2$ is
free to change its velocity $v_i \in \mathbb{R}^2$, and after a short time $\delta > 0$,
it moves to $p_i + \delta \cdot v_i$, if there is no collisions between the agents
(i.e.\ the distance between any pair of agents is at least the sum of their radii).
For a decentralized navigation approach, it runs independently for each individual agent $a_i$,
and based on the observations of the environment, it updates the velocity in order to guide agent $a_i$
to arrive at the given and fixed destination/target $d_i \in \mathbb{R}^2$.
On the measure of the approach's performance, we want all the agents arriving at their destinations/targets
as soon as possible, without causing any collisions.

\paragraph{Our contributions.} We introduced the concept of the \emph{local action cell}
to specify the underlying choices for the selection of the motion to perform,
and proposed two approaches (LAC-Nav and LAC-Learn)
that guarantee to provide the collision-free navigations.
While the LAC-Nav approach simply perform the action
of the largest penalized length (among all choices in the local action cell),
the LAC-Learn approach evaluates the performed actions and adjust
the selection based on an adaptive learning framework.
The experiment results have shown that the proposed approaches perform more efficiently
in the completion time (formally defined in the section of ``Experiments''),
compared to several well studied approaches.

\paragraph{Related works.} The velocity-based collision-free navigation have been
extensively studied in the last two decades.
The idea of \emph{reciprocal velocity obstacles} (RVO, \cite{Berg2008})
was introduced to reduce the problem of calculating the collision-free motion
to solving a low-dimensional linear program,
based on the definition of velocity obstacles \cite{Fiorini1998},
and it was further improved to derive the \emph{optimal reciprocal collision avoidance}
(ORCA, \cite{Berg2011}) framework, which guarantees the optimal reciprocal collision-free motions,
except for some certain conditions with densely packed agents.
While the safety of the final motion is guaranteed by ORCA,
the ALAN \cite{Godoy2015} online learning framework has been proposed
for adapting the preferred motions of multiple agents without the need for
offline training; and the CNav \cite{Godoy2016} is designed to
allow the agents to take the others' preferred motion into account
and adjust accordingly to achieve the better coordination in the crowd environments.
Notice that although the efficiency of CNav has been demonstrated through the experiments,
it requires the the spreading of some private information of the agents, such as
their preferred motions or their targets, which is often a controversial issue in the practical applications.

As the well known Voronoi diagram can be used to divide the working space
into non-intersecting areas, it has been also adopted
for the collision-free path planning with multiple robots \cite{Garrido2006,Bhattacharya2008}.
Inspired by the algorithms for the coverage control of the agents \cite{Pimenta2008},
and a Voronoi-cell-based algorithm \cite{Bandyopadhyay2014} which is introduced to avoid collisions
within a larger probabilistic swarm,
the \emph{buffered Voronoi cell} (BVC, \cite{Zhou2017}) approach
has been proposed to achieve the collision avoidance guarantee for the multiagent navigation,
based on only the information of the positions.
With the up-to-date information of the others' positions,
the agents are restricted to move in the non-intersecting areas,
and thus there should be no collisions.
In \cite{Senbaslar2019}, a trajectory planning algorithm was proposed to
navigate the agents under the higher-order dynamic limits,
in which BVC is used as the low-level strategy to avoid collisions.

\section{The Local Action Cells}
\label{sec:lac}

In this paper, we assume that all the agents in $\mathcal{A}$ have the same radius $r$
for the simplicity of the argument (for the case when the agents have different radii,
the arguments in this paper can be directly extended by substituting the classical Voronoi
diagram with its weighted variant).
Thus for any time and any pair of non-colliding agents $a_i$ and $a_j$,
it always holds that $\|p_{ij}\|_2 \geq 2r$, where $p_{ij}$ stands for $p_j - p_i$.

Recall that in \cite{Zhou2017}, the buffered Voronoi cell of agent $a_i$ is defined as
\[\bar{\mathcal{V}}_i = \left\{p \in \mathbb{R}^2 \mid \left(p - \frac{p_i + p_j}{2}\right) \cdot
\frac{p_{ij}}{\|p_{ij}\|_2} + r \leq 0, \forall j \neq i\right\},\]
which implies a safe velocity domain
\[\mathcal{D}_i = \left\{v \in \mathbb{R}^2| p_i + \delta \cdot v \in \bar{\mathcal{V}}_i \right\},\]
for agent $a_i$ to change and maintain its velocity in order to reach a point in $\bar{\mathcal{V}}_i$,
where $\delta \in \mathbb{R}^+$ is the length of the time interval between two consecutive updates.
Equivalently, domain $\mathcal{D}_i$ can be presented as
\[\mathcal{D}_i = \left\{v \in \mathbb{R}^2 \mid v \cdot u_{ij}
\leq \frac{\|p_{ij}\|_2 - 2r}{2\delta}, \forall j \neq i\right\},\]
where $u_{ij}$ is the unit vector along the same direction with $p_j - p_i$,
i.e.\ $u_{ij} = p_{ij} / \|p_{ij}\|_2$.
Obviously, domain $\mathcal{D}_i$ is the intersection of the half-planes $P_{ij}$'s
for each agent $a_j \neq a_i$, with
\[P_{ij} = \left\{v \in \mathbb{R}^2 \mid v \cdot u_{ij}
\leq \frac{\|p_{ij}\|_2 - 2r}{2\delta} \right\}.\]
Assuming that agent $a_i$ is moving at velocity $v_i$ and agent $a_j$ is moving at velocity $v_j$,
we estimate the colliding risk by calculating
\[\emph{v}_{ij} = \max\left\{0, \frac{\|p_{ij}\|_2 - 2r}{2\delta} - v_j \cdot u_{ij}\right\}\]
and
\[\theta_{ij} = \min\left\{1, \frac{\|p_{ij}\|_2 - 2r}{\emph{v}_{ij} \cdot \tau}\right\},\]
and define the \emph{safe half-plane} $\bar{P}_{ij}$ of agent $i$ according to agent $a_j$ as
a subset of $P_{ij}$
\[\bar{P}_{ij} = \left\{(1 - \lambda + \theta_{ij} \cdot \lambda) \cdot v | v \in P_{ij} \right\},\]
where $\lambda \in [0, 1]$ is the relax factor indicating how much the agent considers
the long-sighted decision, and it is set to $0.5$ through this paper.

\begin{figure}[!htb]
  \centering
  \includegraphics[width=0.48\textwidth]{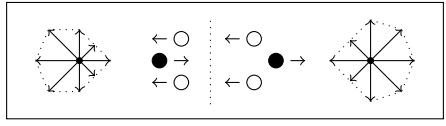}
  \caption{The local action cell of an agent (the black one) moving through two neighbors.}
  \label{fig:lac}
\end{figure}

Now, we are ready to define the \emph{local action cell} (LAC) of agent $a_i$,
denoted by $\mathcal{C}_i$,
as a subset of velocities in the intersection of all the safe half-planes, i.e.\
\begin{eqnarray*}
\mathcal{C}_i &=& \left\{v \in \cap_j \bar{P}_{ij} \mid
\|v\|_2 \leq \min\{\emph{v}_{max}, \frac{\|d_i - p_i\|_2}{\delta}\},\right.\\
&& \left(\rho(v) - \rho(d_i - p_i)\right)~\textrm{mod}~2\pi \in \Delta \bigg\}
\end{eqnarray*}
where $\emph{v}_{max}$ indicates the maximum moving speed,
$d_i$ is the destination/target of agent $a_i$,
$\rho(\cdot)$ denotes the angle (in radians) of
the clockwise rotation of the argument vector
to align with the positive direction of the $x$-axis,
and $\Delta$ is a set of candidate angles which is defined by
\[\Delta = \left\{k \cdot \frac{\pi}{4} \mid k \in \mathbb{Z}, 0 \leq k < 8\right\}, \]
through this paper. (See Figure~\ref{fig:lac} for an illustration of the local action cell
of an agent moving through two neighbors.)

\section{Collision-Free Navigation}

In this section, we introduce a distributed approach, named LAC-Nav,
for the collision-free navigation with multiple agents.
As shown in Algorithm~\ref{algo:lac-nav}, the approach is straight forward
with the following steps executed in loops:
for each agent $a_i$, calculate the current local action cell;
and then select a proper velocity from the cell.

Algorithm~\ref{algo:lac} follows the definition of the local action cell and
describes the calculation details; Algorithm~\ref{algo:select_vel} shows
how the new velocity is selected: Given the current local action cell $\mathcal{C}_i$,
each velocity $v \in \mathcal{C}_i$ is at first evaluated according to the
penalized length $\zeta_v \cdot \|v\|_2$, where $\zeta_v$ is the factor
that is initialized as $0 < \zeta \leq 1$ and
decreased exponentially on the angle between $v$ and the direction of $d_i - p_i$.
Finally, the velocity of the maximum penalized length is returned as the result.

\begin{algorithm}[!htb]
\caption{LAC-Nav$(a_i)$: The LAC-based navigation algorithm running on agent $a_i$.}
\label{algo:lac-nav}
\nl \While{$a_i$ is not at the destination}{
  \nl $C_i :=$ LAC$(a_i)$;\\
  \nl $v^{new}_i :=$ SelectVel$(C_i)$;\\
  \nl agent $a_i$ moves at velocity $v^{new}_i$;\\
}
\end{algorithm}

\begin{algorithm}[!htb]
\caption{LAC$(a_i)$: Calculate the current local action cell of agent $a_i$.}
\label{algo:lac}
\nl $\emph{v}^{max}_i := \min\{\emph{v}_{max}, \frac{\|d_i - p_i\|_2}{\delta}\}$;\\
\nl $v^0_i := \emph{v}^{max}_i \cdot \frac{d_i - p_i}{\|d_i - p_i\|_2}$;\\
\nl \For{$k = 1$ to $7$}{
    \nl calculate $v^k_i$ such that $\|v^k_i\|_2 = \emph{v}^{max}_i$
    and $\rho(v^k_i) = \left(\rho(v^0_i) + k \cdot \pi/4\right)$ mod $2\pi$;\\
}
\nl $C_i := \{v^k_i | k \in \mathbb{Z}^*,  0 \leq k \leq 7\}$;\\
\nl \For{agent $a_j$ with $j \neq i$}{
    \nl calculate the safe half-plane $\bar{P}_{ij}$;\\
    \nl \For{$v \in C_i$}{
        \nl $\Gamma_v := \{\theta \cdot v \mid \theta \cdot v \in \bar{P}_{ij},
        0 \leq \theta \leq 1\}$;\\
        \nl $v := \arg\max_{u \in \Gamma_v} \|u\|_2$;\\
    }
}
~\\
\nl \textbf{Return}: $C_i$;\\
\end{algorithm}

\begin{algorithm}[!htb]
\caption{SelectVel$(C_i)$: Select a velocity inside cell $C_i$ as the new velocity to move at.}
\label{algo:select_vel}
\nl \For{$v \in C_i$}{
\nl $\alpha_v := \left(\rho(v) - \rho(d_i - p_i)\right)$ mod $2\pi$;\\
\nl $\zeta_v := \zeta^{\frac{4\alpha_v}{\pi}}$;\\
\nl $w_v := \zeta_v \cdot \|v\|_2$;\\
}
\nl $v^{new}_i := \arg\max_{v\in C_i} w_v$
~\\
\nl \textbf{Return}: $v^{new}_i$;\\
\end{algorithm}

While calculating the local action cells, it is not necessary to consider
all the agents in the environment.
When the distance between agent $a_i$ and agent $a_j$ is at least
$\ell := 2 \cdot \emph{v}_{max} \cdot \tau + 2 \cdot r$,
it holds directly that $\theta_{ij} = 1$ and $\theta_{ji} = 1$.
Thus the corresponding safe half-planes can be ignored in the calculation
of the agents' local action cells, which implies it is sufficient
to consider only the neighbors within distance $\ell$.

\paragraph{Processing complexity.}
When considering only the agents within a distance $\ell$,
the number of an agent's neighbors is at most $3 \cdot \ell^2 / r^2$,
since there is no overlap between the neighbors and for each of them,
at least $1/3$ of the body is covered by the disk of radius $\ell$.
Consequently, the loop of Lines $6-10$ is executed for a constant time
within one step of update of an individual agent.
Thus, the processing complexity of LAC is determined
by the efficiency to detect the neighbors in the specified range.
In the simulations, the neighbors can be efficiently derived
through querying in a KD-Tree that maintains all the positions,
and in more practical cases, the neighbor detection is often
executed in a parallel process, and it can be assume that
the required information is always ready when it is needed.

\paragraph{Learning with LAC.} In LAC-Nav, the new velocity is selected
according to the penalized length, which can be roughly seen as an estimate
of the traveling distance of the next move. On the other hand, it is also common
to evaluate the performed actions and record the results,
which also provides the information that may be useful
for making decisions in the future.
In the case when a specific behavior should perform well for a period of time,
selecting the action of the best known evaluation should be more
promising than trying based on the estimates only.
Generally, the evaluations are learned as the agent keeps running
in the ``sense-evaluate-act'' cycles.

Following the ALAN learning framework \cite{Godoy2015},
we propose the LAC-Learn approach, in which the reward of the latest performed
action is defined as the summation of the penalized lengths of the velocities
in the resulting local action cell.
Notice that by this definition, the reward naturally incorporates
the considerations of the goal-oriented performance and the politeness performance,
which are treated as two separate components in ALAN.
In fact, the lengths of the velocities approaching to the destination
reflect how efficient the performed action is for getting the agent closer to the goal;
and the lengths of velocities in the local action cell as a whole reflects
the efficiency in avoiding the crowding situations.
In spite of the definition of the action reward,
LAC-Learn selects the new velocity in a different way from the one used in ALAN.
With LAC-Learn, the selected new velocity is the one corresponding to the action
that maximizes a linear combination of the reward and the penalized velocity length.

\begin{algorithm}[!htb]
\caption{LAC-Learn$(a_i)$: Navigation algorithm of agent $a_i$ while learning with the local action cells.}
\label{algo:lac-learn}
\nl \While{$a_i$ is not at the destination}{
  \nl $C_i :=$ LAC$(a_i)$;\\
  \nl $W_i :=$ CalcWeights$(C_i)$;\\
  \nl $R_i :=$ UpdateReward$(\alpha_i, W_i, R_i)$;\\
  \nl $S_i :=$ UpdateWUCB$(\alpha_i, R_i, S_i)$;\\
  \nl $\alpha_i :=$ SelectAct$(\alpha_i, W_i, R_i, S_i)$;\\
  \nl $v^{new}_i := C_i[\alpha_i]$;\\
  \nl agent $a_i$ moves at velocity $v^{new}_i$;\\
}
\end{algorithm}

\begin{algorithm}[!htb]
\caption{SelectAct$(\alpha_i, W_i, R_i, S_i)$: Select the action for agent $a_i$ to perform.}
\label{algo:select_act}
\nl $\epsilon_i := 0$;\\
\nl $\alpha :=$ Null;\\
\nl \If{$\alpha_i = 0$}{
    \nl \If{$W_i[0] \geq \eta \cdot \min\{\emph{v}_{max}, \frac{\|d_i - p_i\|_2}{\delta}\}$}{
        \nl $\alpha := 0$;\\
        \nl $\epsilon_i := 0$;\\
    }
    \nl \Else {
        \nl $\epsilon_i := \min\{1, \epsilon_i + \beta\}$;\\
    }
}
\nl \If{$\alpha =$ \emph{Null}}{
    \nl take $s$ from $[0, 1]$ uniformly at random;\\
    \nl \If{$s < 1 - \epsilon_i$}{
    \nl $\alpha := \arg\max_{\alpha \in \Delta}\left((1 - \gamma) \cdot R_i[\alpha] + \gamma \cdot W_i[\alpha]\right)$;\\
    }
    \nl \Else {
    \nl $\alpha := \arg\max_{\alpha \in \Delta} S_i[\alpha]$;\\
    }
}
~\\
\nl \textbf{Return}: $\alpha$;\\
\end{algorithm}

Inside an execution cycle of some agent $a_i$, after the local action cell
is calculated by LAC (Algorithm~\ref{algo:lac}),
the penalized length of each velocity in $C_i$ is calculated as
what has been done in Line~$2-4$ of SelectVel (Algorithms~\ref{algo:select_vel}),
and saved in a set $W_i$. In UpdateReward,
the reward of the last performed action is updated to
the sum of all weights in $W_i$, as mentioned earlier.

Notice that although the velocities given by $C_i$ may vary from step to step,
in the local view, they can always be interpreted as the actions corresponding to
the angles specified in $\Delta$.
For example, without considering the variation of the length,
the velocity pointing to the destination can always be interpreted
as the action corresponding to angle $0 \in \Delta$.
For an action/angle $\alpha \in \Delta$,
we use $C_i[\alpha]$ to denote the velocity $v \in C_i$
such that $(\rho(v) - \rho(d_i - p_i))~\textrm{mod}~2\pi = \alpha$.

Following the ALAN learning framework, we calculate (by UpdateWUCB)
and maintain (in $S_i$) the upper confidence bound within a moving time window
(i.e.\ a sequence of consecutive time steps),
which is used when the agent explores in the action space.
As defined in \cite{Godoy2015}, the wUCB score of action $\alpha$
during the last $T \in \mathbb{Z}^{+}$ steps is defined by
\[\textrm{wUCB}(\alpha) := \bar{R}_i(\alpha) + \sqrt{\frac{2\ln(\nu)}{\nu_\alpha}},\]
where $\bar{R}_i(\alpha)$ is the average reward of action $\alpha$,
$\nu_\alpha$ denotes the number of times action $\alpha$ has been chosen,
and $\nu$ denotes the total number of performed actions,
all with respect to the moving time window.

Similar to the context-aware action selection approach proposed in \cite{Godoy2015},
SelectAct (Algorithm~\ref{algo:select_act}) decides
with the ``win-stay, lose-shift'' strategy and
the adaptive $\epsilon$-greedy strategy in which the wUCB suggested action
is chosen for the exploration.

When the agent is in the winning state (i.e.\ the goal-oriented action
$\alpha_i = 0$ is performed in the last update and is still a good
choice in the sense that the corresponding velocity is little constrained),
it is natural to keep forwarding to the goal.
Otherwise, if the agent is in the losing state,
it performs the $\epsilon$-greedy strategy
to exploit on the action that maximizes a linear combination
of the action reward and the penalized length of the corresponding velocity.
With a small and adaptively adjusted probability,
the agent explores and performs the action that maximizes the wUCB score.

the hyper-parameters $\eta \in [0, 1]$ (Line~$4$ in Algorithm~\ref{algo:select_act})
and $\gamma \in [0, 1]$ (Line~$12$ in Algorithm~\ref{algo:select_act})
are determined depends on the scenarios.

\section{Experiments}

In this section, we present the results of running experiments
with LAC-Nav and LAC-Learn, on a computer of
$7$ Intel Core i7-6700 CPU ($3.40$ GHz) processors.
The simulations are implemented in Python $3.5$,
while the update processes of individual agents have been speeded up
by applying the multitasking scheme.
For one second, there are $100$ updates performed
for each agent, and therefore we set $\delta := 0.01$
in the implementation of LAC-Nav and LAC-Learn.

\begin{figure}[!htb]
  \centering
  \subfigure[Reflection: start]
  {\frame{\includegraphics[width=0.25\textwidth]{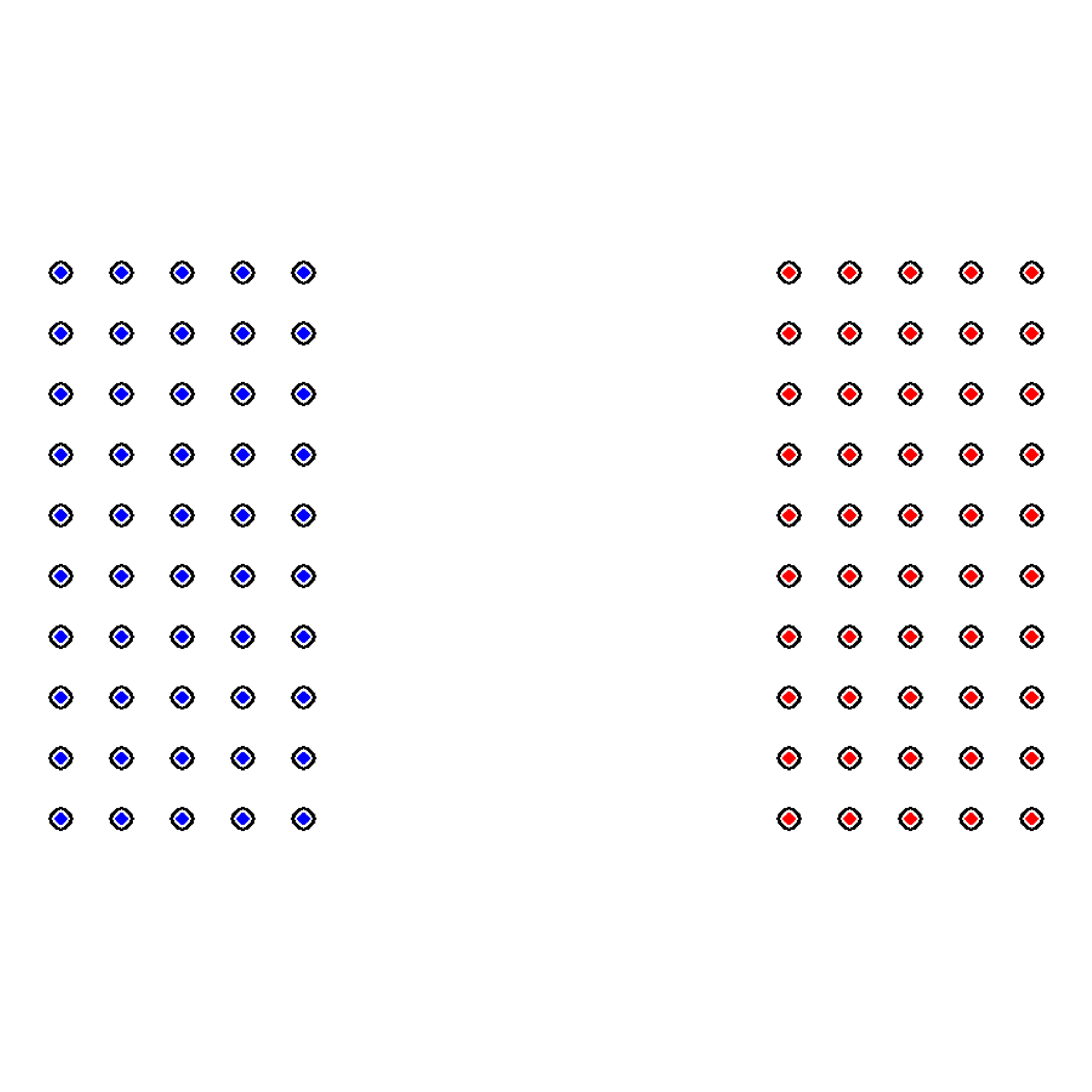}\label{fig:against_starts}}}
  \hfill
  \subfigure[Circle: start]
  {\frame{\includegraphics[width=0.25\textwidth]{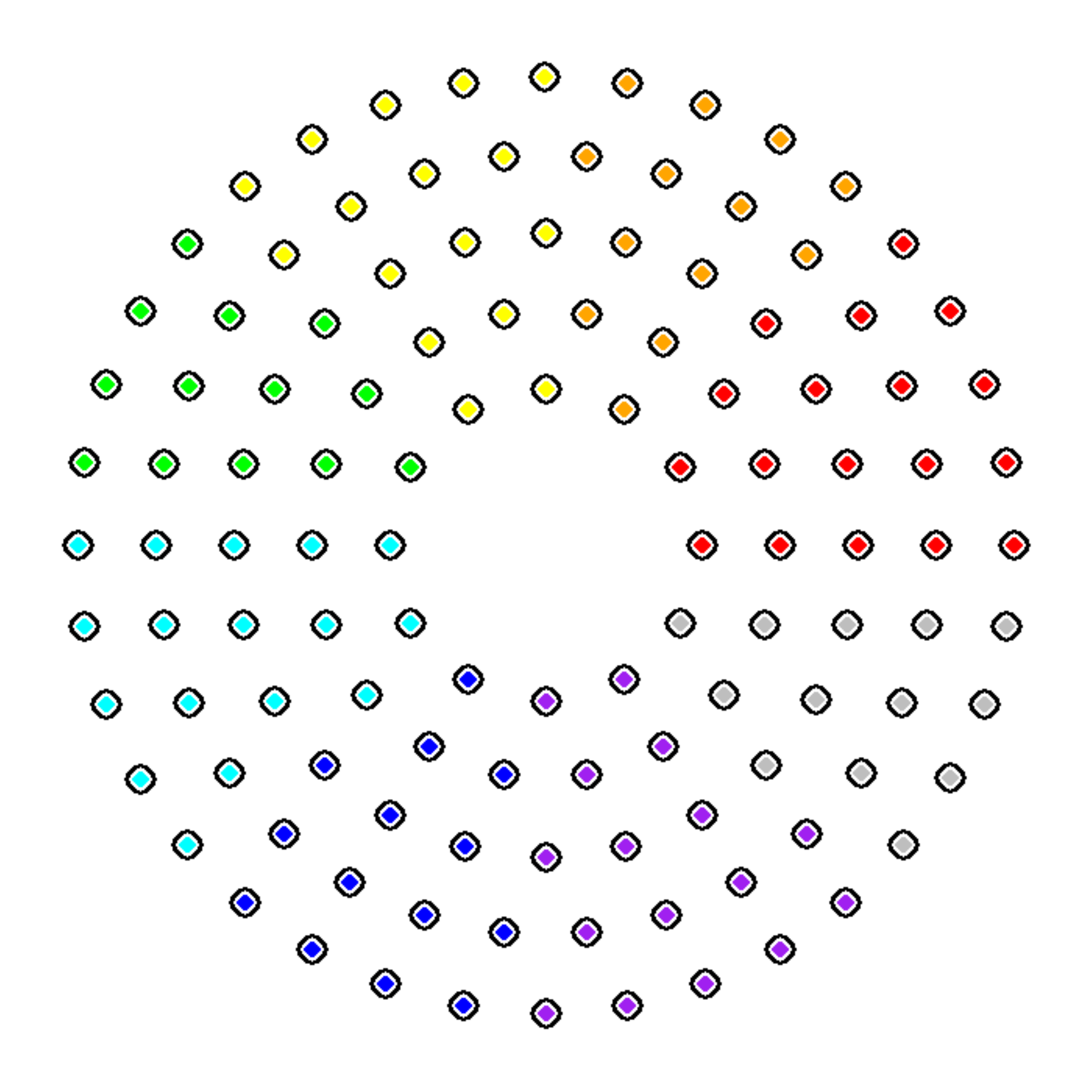}\label{fig:circle_starts}}}
  \hfill
  \subfigure[Crowd: start]
  {\frame{\includegraphics[width=0.25\textwidth]{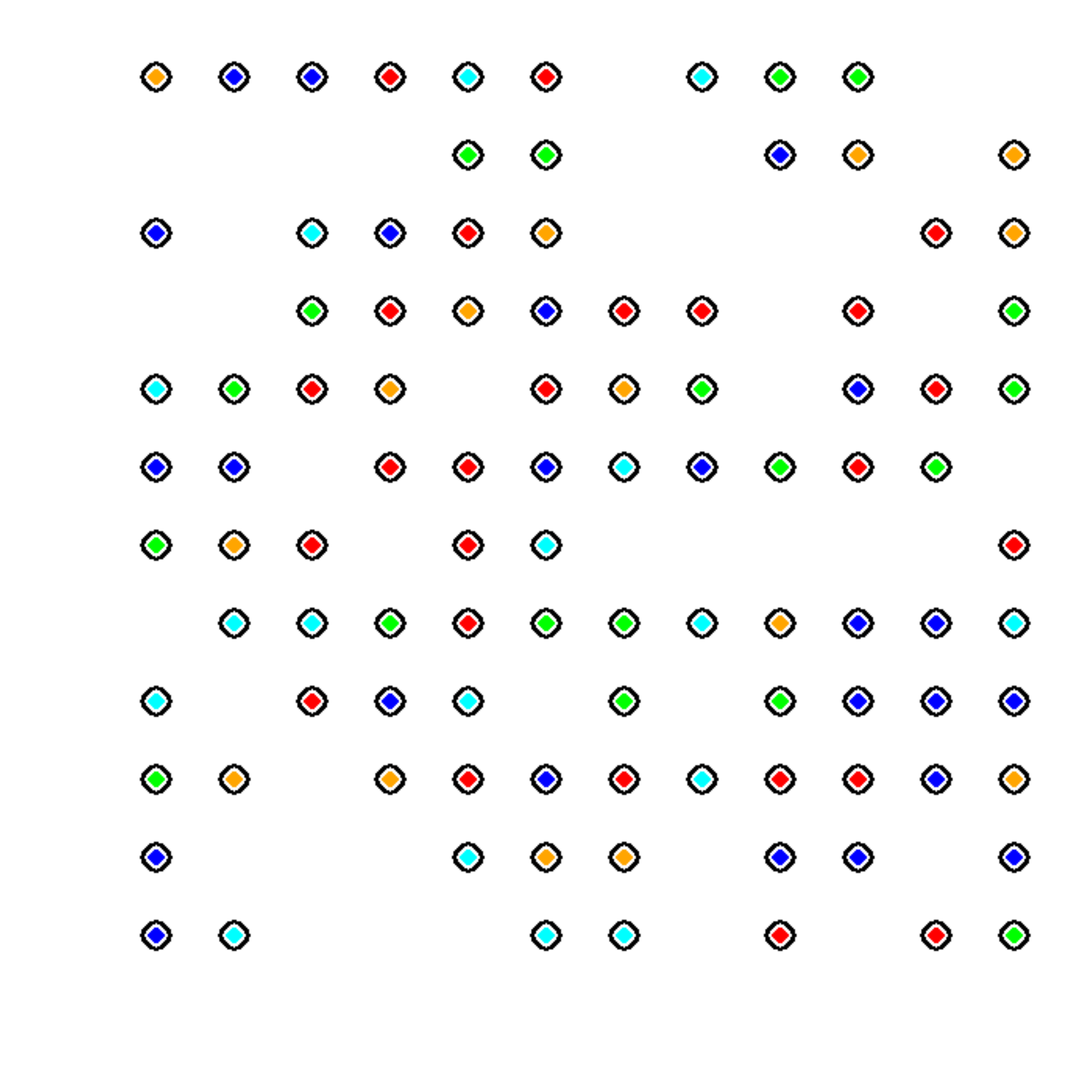}\label{fig:crowd_starts}}}
  \subfigure[Reflection: target]
  {\frame{\includegraphics[width=0.25\textwidth]{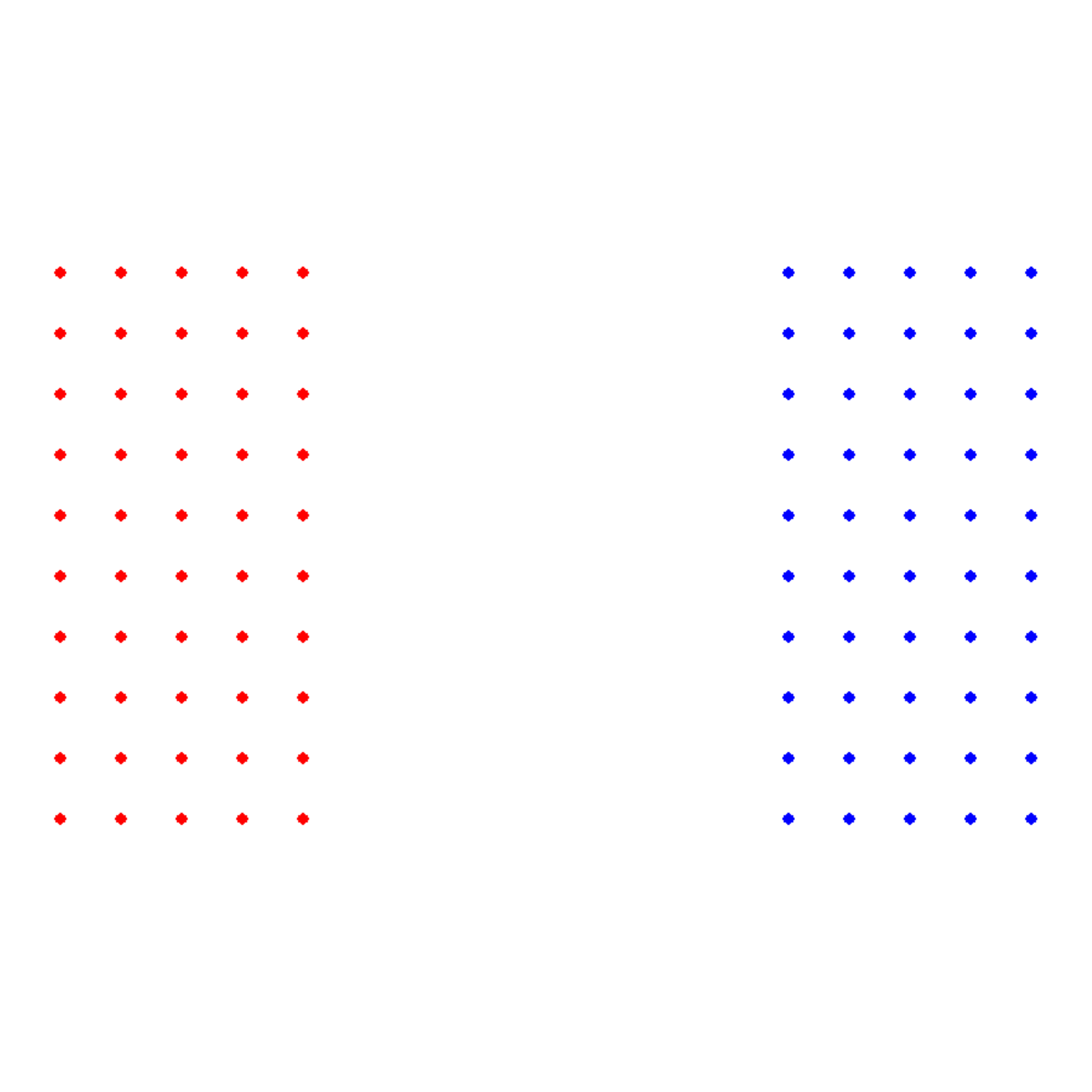}\label{fig:against_targets}}}
  \hfill
  \subfigure[Circle: target]
  {\frame{\includegraphics[width=0.25\textwidth]{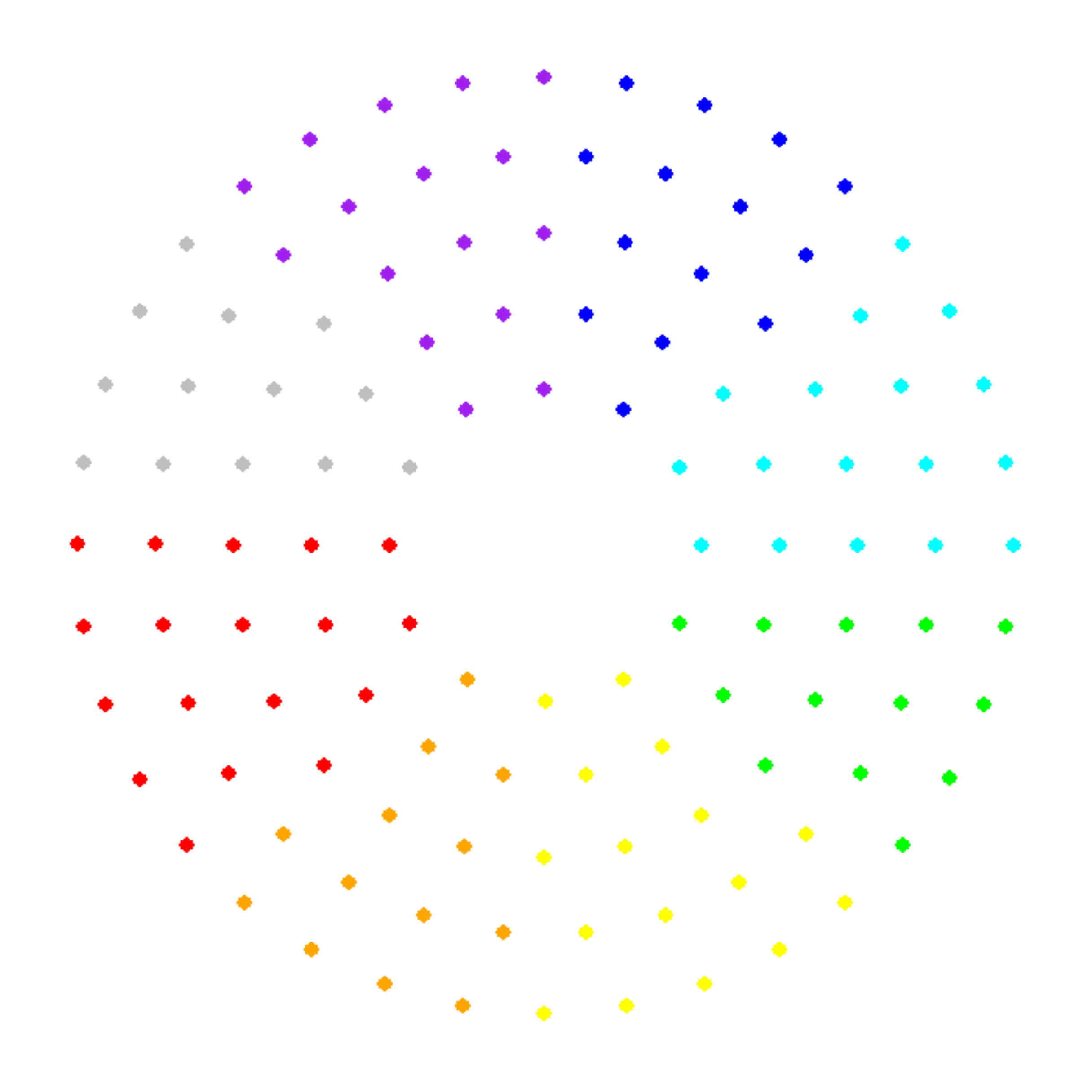}\label{fig:circle_targets}}}
  \hfill
  \subfigure[Crowd: target]
  {\frame{\includegraphics[width=0.25\textwidth]{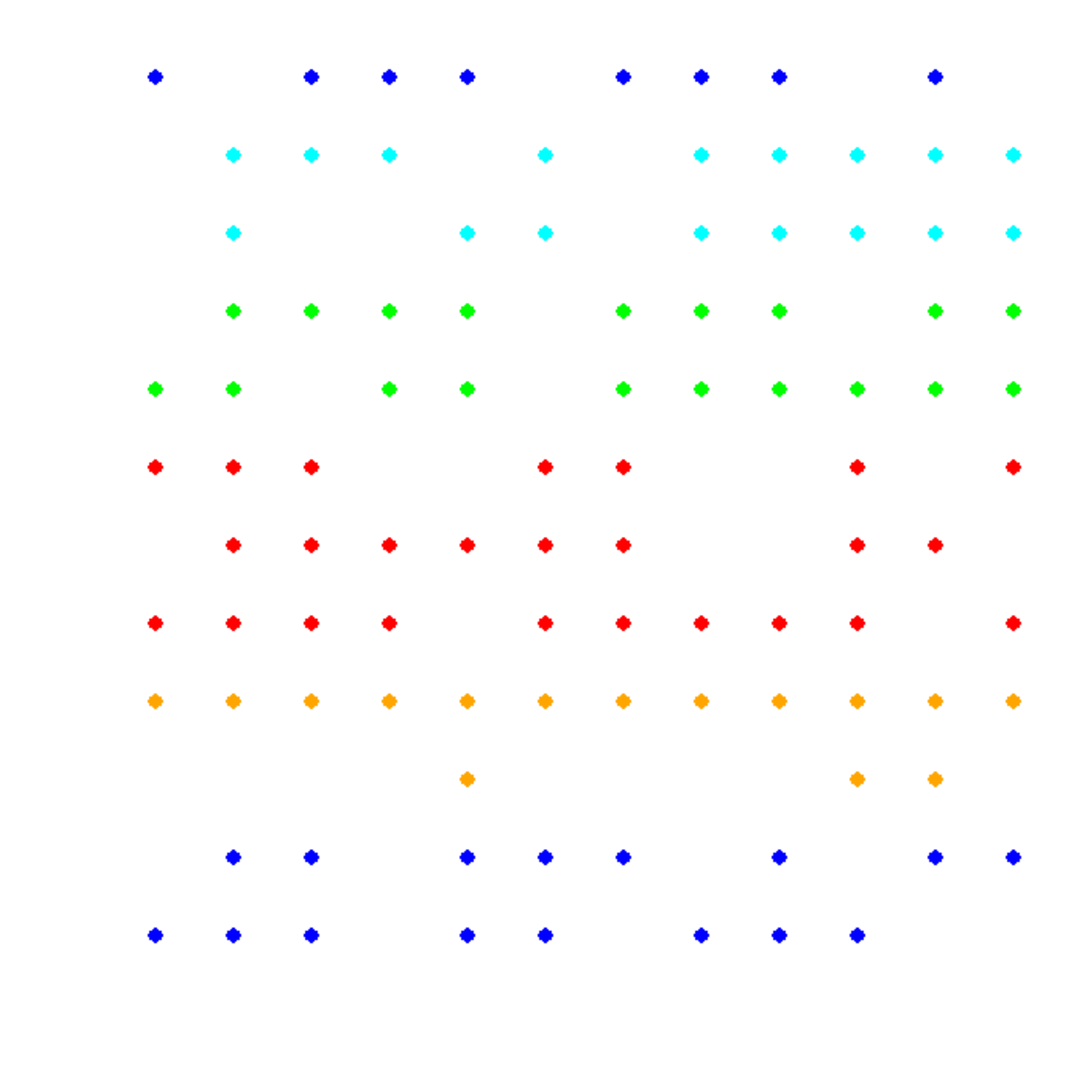}\label{fig:crowd_targets}}}
  \caption{Experiment scenarios.}
  \label{fig:scenarios}
\end{figure}

\paragraph{Scenarios.}
For the experiments, we considered three scenarios (Figure~\ref{fig:scenarios}):
the \emph{reflection} scenario, the \emph{circle} scenario and the \emph{crowd} scenario,
where
\begin{itemize}
  \item in the reflection scenario, two groups of agents start from the left side and right side
  of the area, respectively (Figure~\ref{fig:against_starts}).
  For each agent, the target is the position on the other side that is symmetric to its start position
  (Figure~\ref{fig:against_targets}).
  Through navigating the agents to the target positions,
  the picture of the starting configuration is reflected.
  \item in the circle scenario, the agents start in layers of circles (Figure~\ref{fig:circle_starts}),
  and each agent targets the antipodal position (Figure~\ref{fig:circle_targets}).
  That is, the picture of starting configuration is going to be ``rotated''
  by half of a circle, around the origin/center.
  \item in the crowd scenario, the start positions (Figure~\ref{fig:crowd_starts})
  and target positions (Figure~\ref{fig:crowd_targets}) are
  randomly picked from a small area.
\end{itemize}

For each of these three scenarios, we compare the performances of
LAC-Nav and LAC-Learn, with the performances
of approaches including BVC \cite{Zhou2017}, CNav \cite{Godoy2016},
ALAN \cite{Godoy2015} and ORCA \cite{Berg2011}.

In this work, we consider two measurements: the \emph{completion time}
and the \emph{average detour-distance ratio},
as the evaluation of the algorithm's performance for the multiagent navigation tasks.
\begin{itemize}
  \item the completion time of running a navigation algorithm is defined
  as the time (in seconds) when the last agent arrives at its target,
  assuming all the agents start from time $0$;
  \item the average detour-distance ratio
  is defined as the average of the ratios between the actual travel distance and the optimal
  travel distance (i.e.\ the length of the straight line from the start
  position to the target position), over all the agents.
  \item the average detour-time ratio
  is defined as the average of the ratios between the actual travel time and the optimal
  travel time (i.e.\ the time of moving in a straight line from the start position
  to the target position, at the maximum speed), over all the agents.
\end{itemize}

While the completion time justifies the algorithm's global performance
on finishing the navigation tasks, by investigating the detour-distance/time ratio,
it provides a view on the variance of the individual agent's behavior
with different algorithms.

In the experiments for all scenarios, the agent's radius is uniformly
set as $r = 10$, and the maximum moving speed is set as $\emph{v}_{max} = 50$.
In addition, as mentioned in the beginning of this section,
within each second there are $100$ updates performed
for each of the agents, which implies the
that the time interval between two consecutive updates is $0.01$,
i.e.\ $\delta = 0.01$ in all the experiments.

Recall that when calculating the local action cells,
the hyper-parameter $\tau > 0$ is needed to locate the safe half-planes.
Through all experiments involving the local action cells, we set $\tau = 0.05$.
In addition, the penalty factor $\zeta$ is also needed in the calculation
of the local action cells (thus it is required when running LAC-Nav
and LAC-Learn). Through the experiments, we set $\zeta = 0.95$,
with which the goal-orthogonal action (with angle $\pi/2$ from the
direction pointing to the goal) is penalized by $0.9025$, and
the goal-opposite action (along the direction leaving the goal)
is penalized by about $0.8145$.

For LAC-Learn, we set the mixing factor (Line~$12$ of Algorithm~\ref{algo:select_act})
as $\gamma = 0.75$ for the reflection scenario and the circle scenario,
and $\gamma = 0.95$ for the crowd scenario.
and the length of the moving time window (for the calculation of wUCB)
as $T = 8$, which is the minimum choice as
there are $8$ actions in the used $\Delta$.
Furthermore, the incremental step (for adjusting the exploration probability)
is set as $\beta = 0.1$ (Line~$8$ of Algorithm~\ref{algo:select_act}).

For ORCA, the collision-free time window is set as $\tau = 0.02$,
i.e.\ twice of the update interval's length.
Recall that with CNav and ALAN, ORCA is also called
to make sure the performed velocity is collision-free,
where the time window are also set as $\tau = 0.02$.
Notice that the time window $\tau > 0$ in ORCA has different meaning
from the hyper-parameter using the same symbol in the calculation
of the local action cells, even though they are both related
to the avoidance of the potential collisions.

For CNav, the hyper-parameter for mixing the goal-oriented reward
and the constrained-reduction reward is set as $\gamma = 0.5$
for the reflection scenario, and $\gamma = 0.9$ for the circle
scenario and the crowd scenario;
the number of constrained neighbors of which the action's effect
is estimated is set as $k = 3$;
the number of the neighbor-based actions is set as $s = 3$.

For ALAN, the hyper-parameter for mixing the goal-oriented reward
and the politeness reward is set as $\gamma = 0.5$;
the length of the moving time window for the calculation of wUCB  is set as $T = 5$;
and the incremental step for adjusting the exploration probability
is set as $\beta = 0.1$.

\begin{figure}[!htb]
  \centering
  \includegraphics[width=0.75\textwidth]{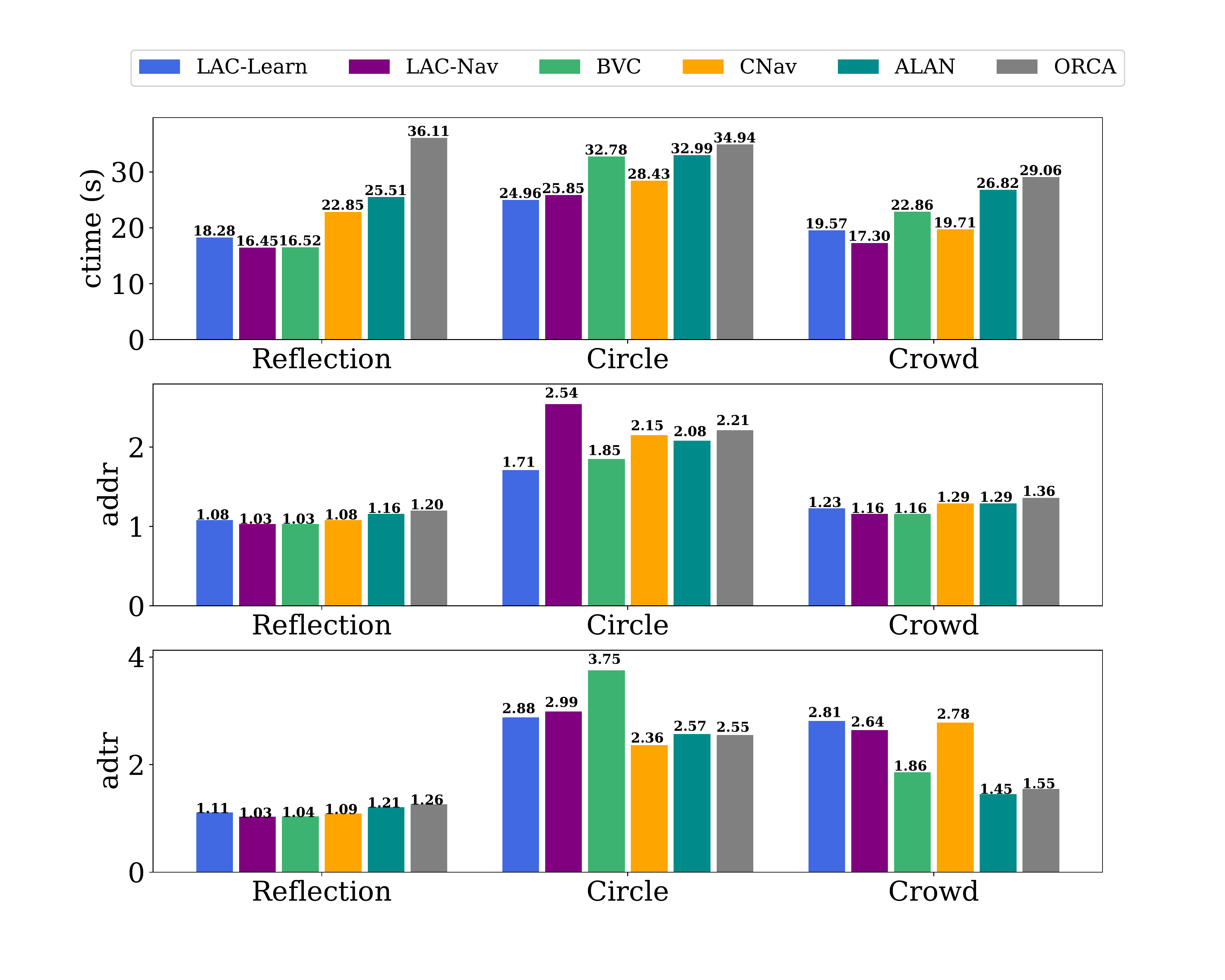}
  \caption{Experiment results, where \emph{ctime} (s) stands
  for the completion time (in seconds); \emph{addr} stands
  for the average detour-distance ratio; and \emph{adtr} stands
  for the average detour-time ratio.}
  \label{fig:results}
\end{figure}

\paragraph{Results.} In Figure~\ref{fig:results}, it shows the experiment results for
\begin{itemize}
  \item the reflection scenario of $100$ agents ($50$ agents on each side);
  \item the circle scenario of $120$ agents (in $5$ circles around the same center point);
  \item the crowd scenario of $100$ agents located in the area of size $600 \times 600$.
\end{itemize}
Overall, LAC-Nav and LAC-Learn outperform
almost all the other approaches in the completion time.
The only exception is in the reflection scenario, BVC
has shown the advantages and it completes earlier than LAC-Learn.

\begin{figure}[!htb]
\centering
  \subfigure[LAC-Nav]
  {\frame{\includegraphics[width=0.25\textwidth]{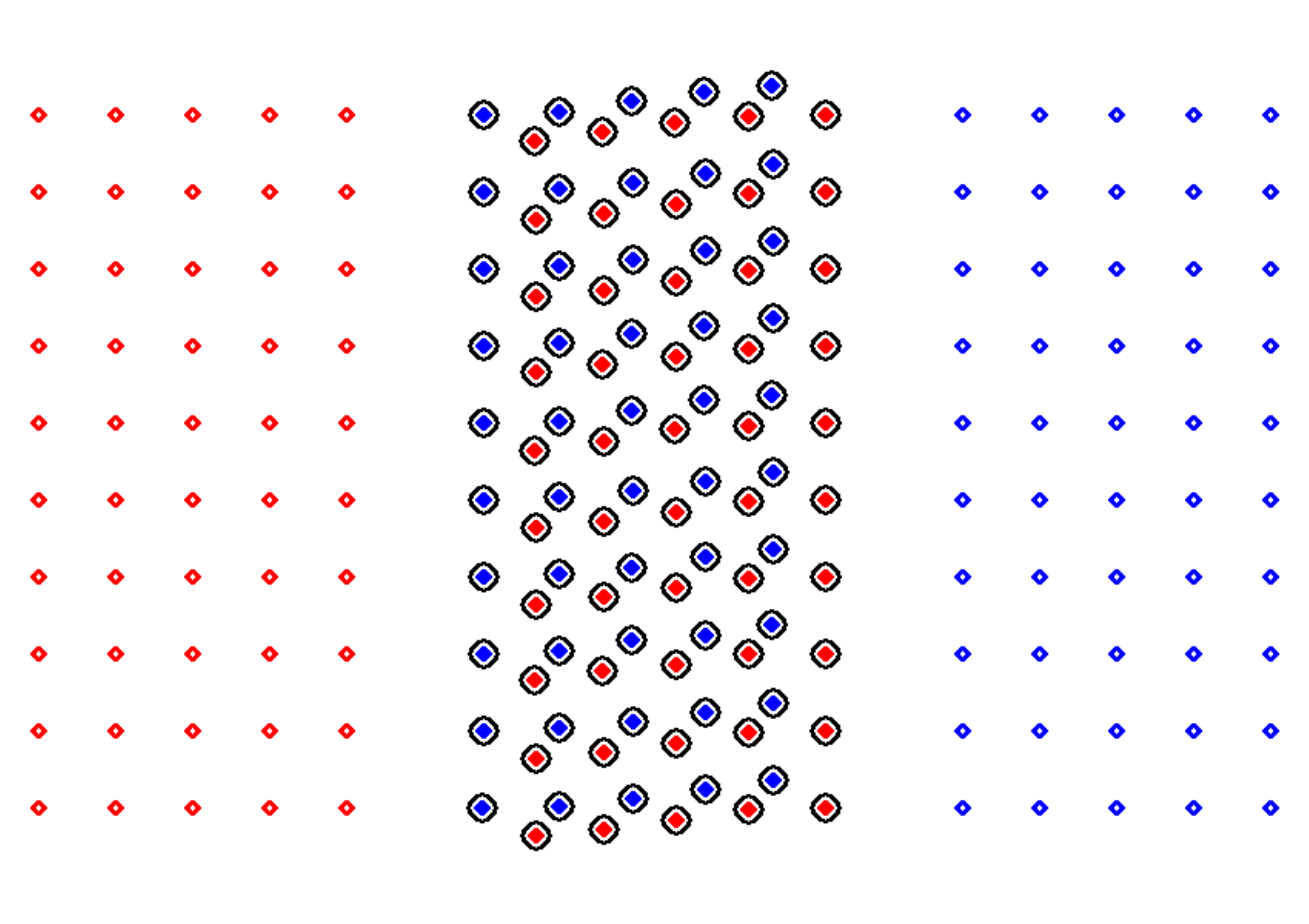}}}
  \hfill
  \subfigure[CNav]
  {\frame{\includegraphics[width=0.25\textwidth]{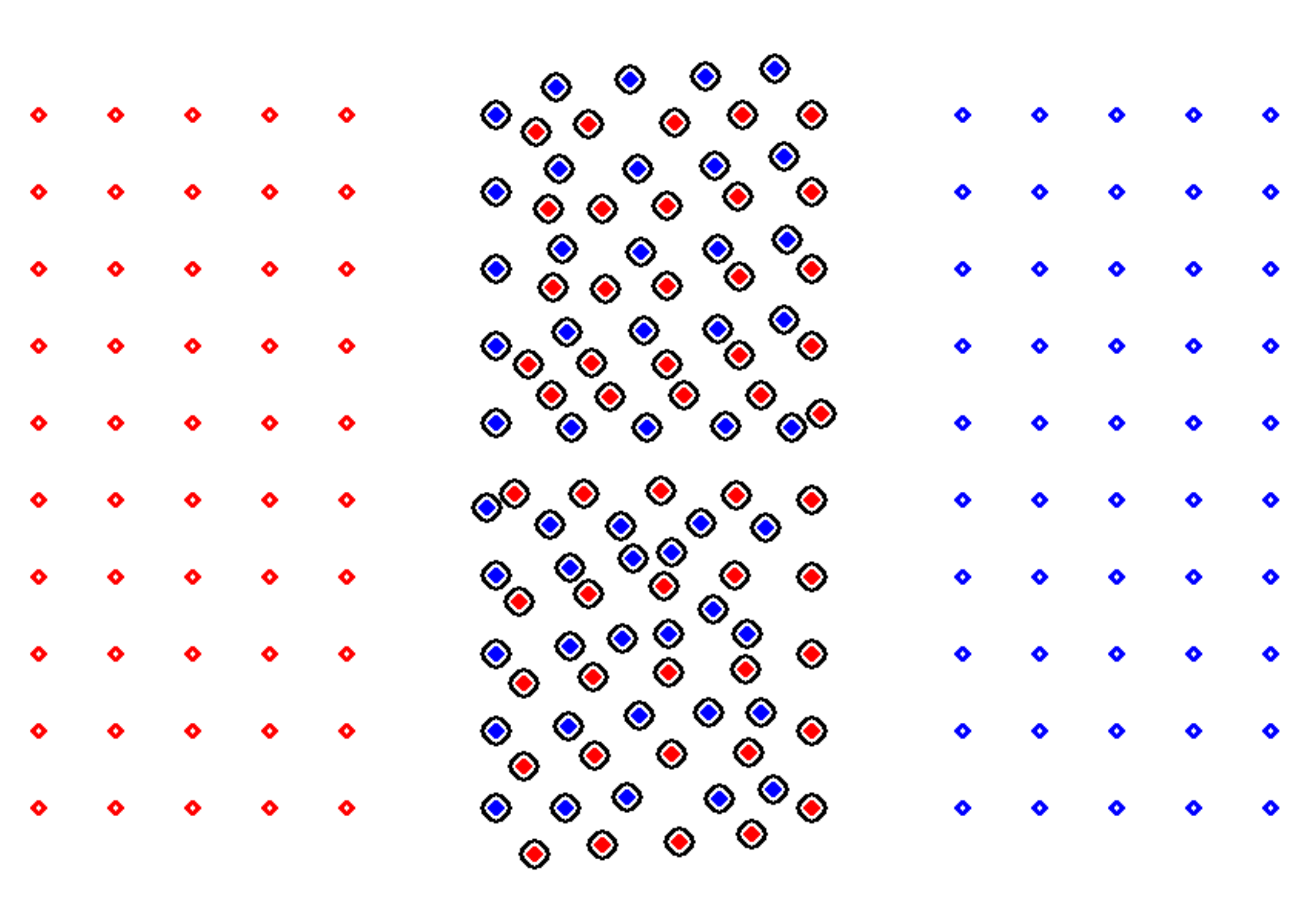}}}
  \hfill
  \subfigure[ORCA]
  {\frame{\includegraphics[width=0.25\textwidth]{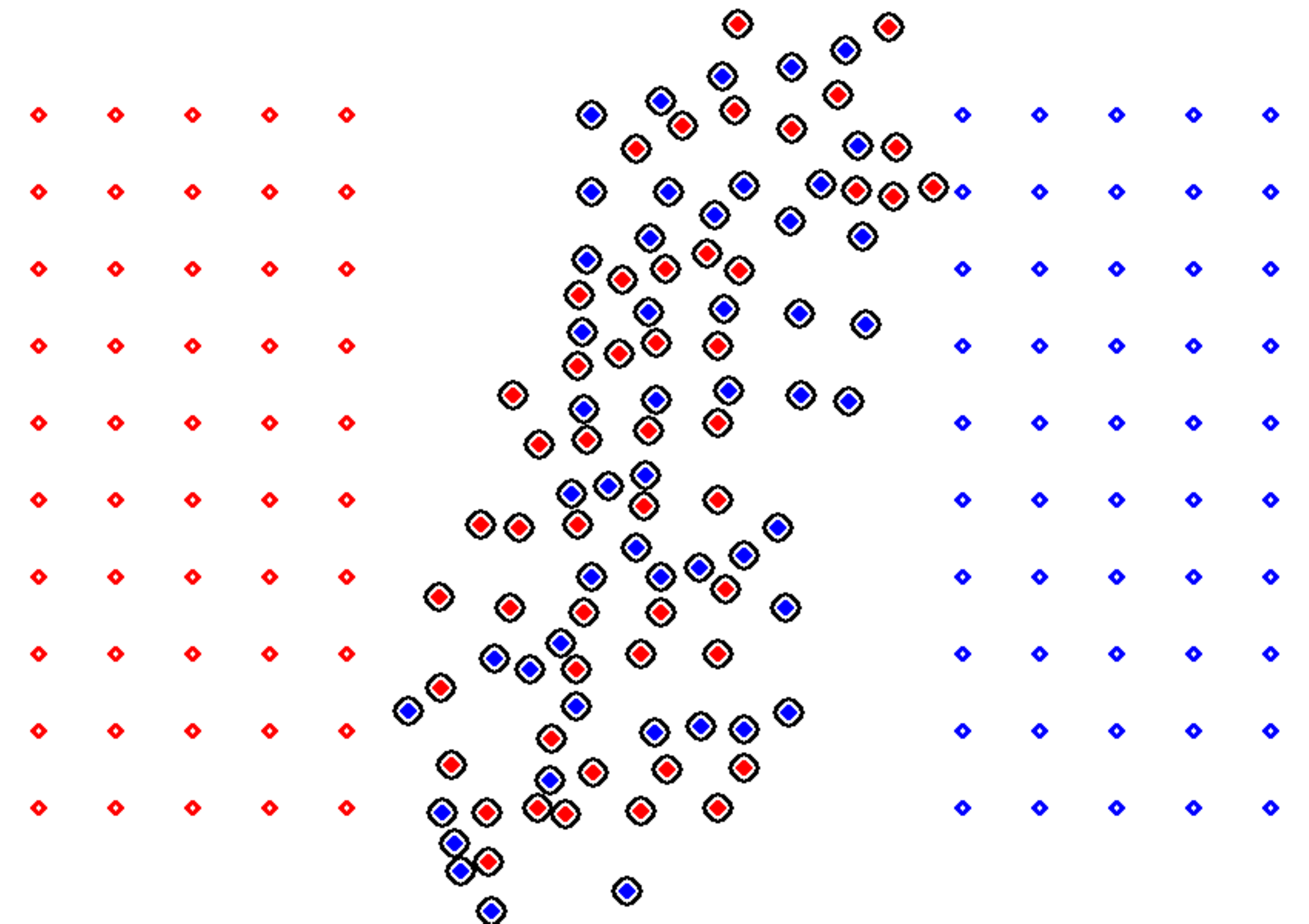}}}
  \caption{The antagonistic agents pass by each other in the reflection scenario,
  where the points of black outline and colored (red/blue) inside are the current positions of the agents,
  and the simply colored (red/blue) points are the target positions of the agents.}
  \label{fig:reflection_pass}
\end{figure}

In general, the efficiency of the LAC based approaches is due to
the fact that it considers both of the task to arrive at the target and
the intension to move as much as possible in every step.
The later consideration prevents the agent from the non-necessary halting
before it arrives at the target.
By maintaining and comparing the penalized lengths of all the candidate actions
(according to $\Delta$), even though the agent still has the change to move
directly towards the target, it detours as long as there is an other action that
provides a better moving (penalized) velocity.
As shown in the reflection scenario, this kind of active detouring
results in a more fluent navigation as the agents (of the antagonistic moving directions)
pass by each other (Figure~\ref{fig:reflection_pass}).

\begin{figure}[!htb]
\centering
  \subfigure[LAC-Nav]
  {\frame{\includegraphics[width=0.25\textwidth]{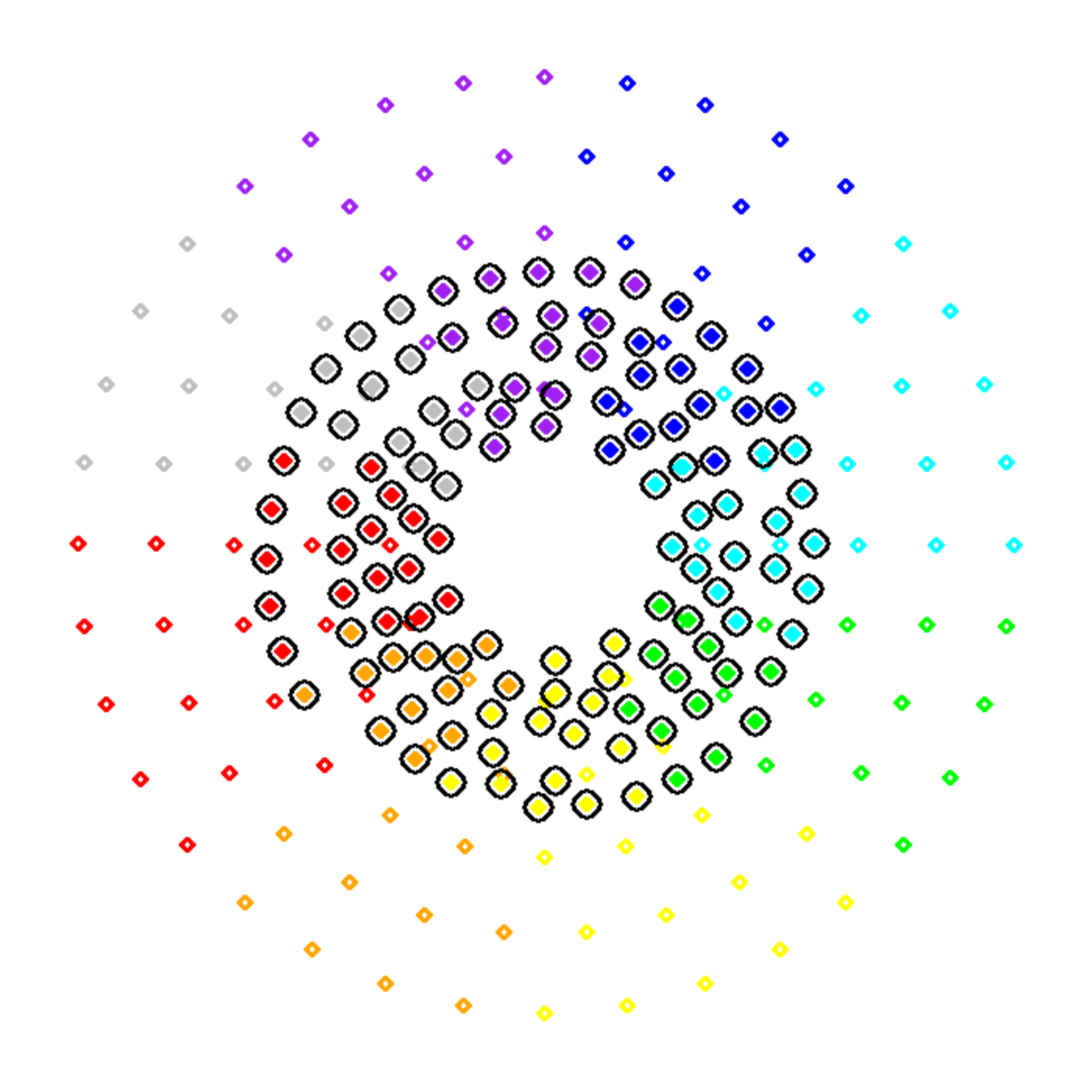}}}
  \hfill
  \subfigure[CNav]
  {\frame{\includegraphics[width=0.25\textwidth]{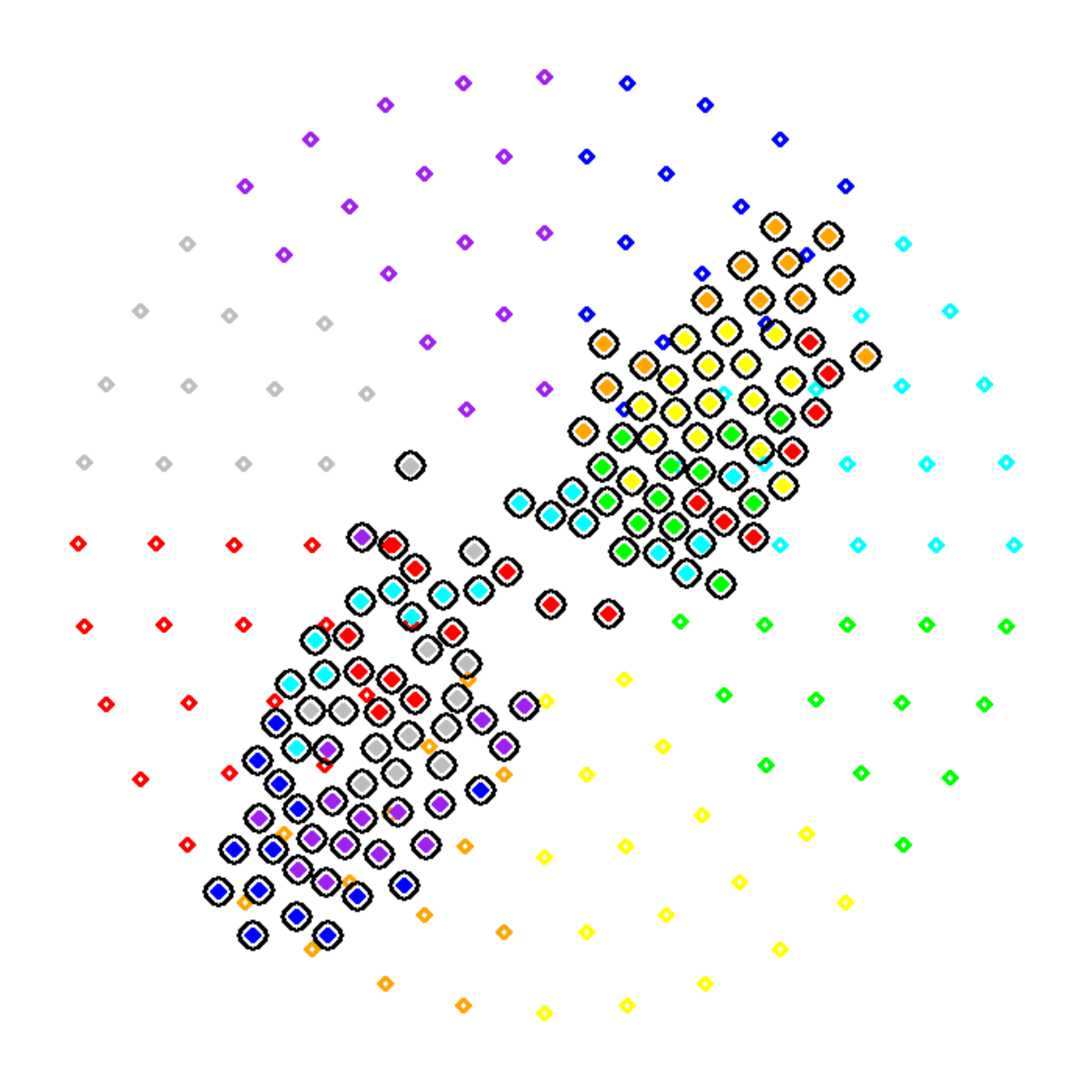}}}
  \hfill
  \subfigure[ORCA]
  {\frame{\includegraphics[width=0.25\textwidth]{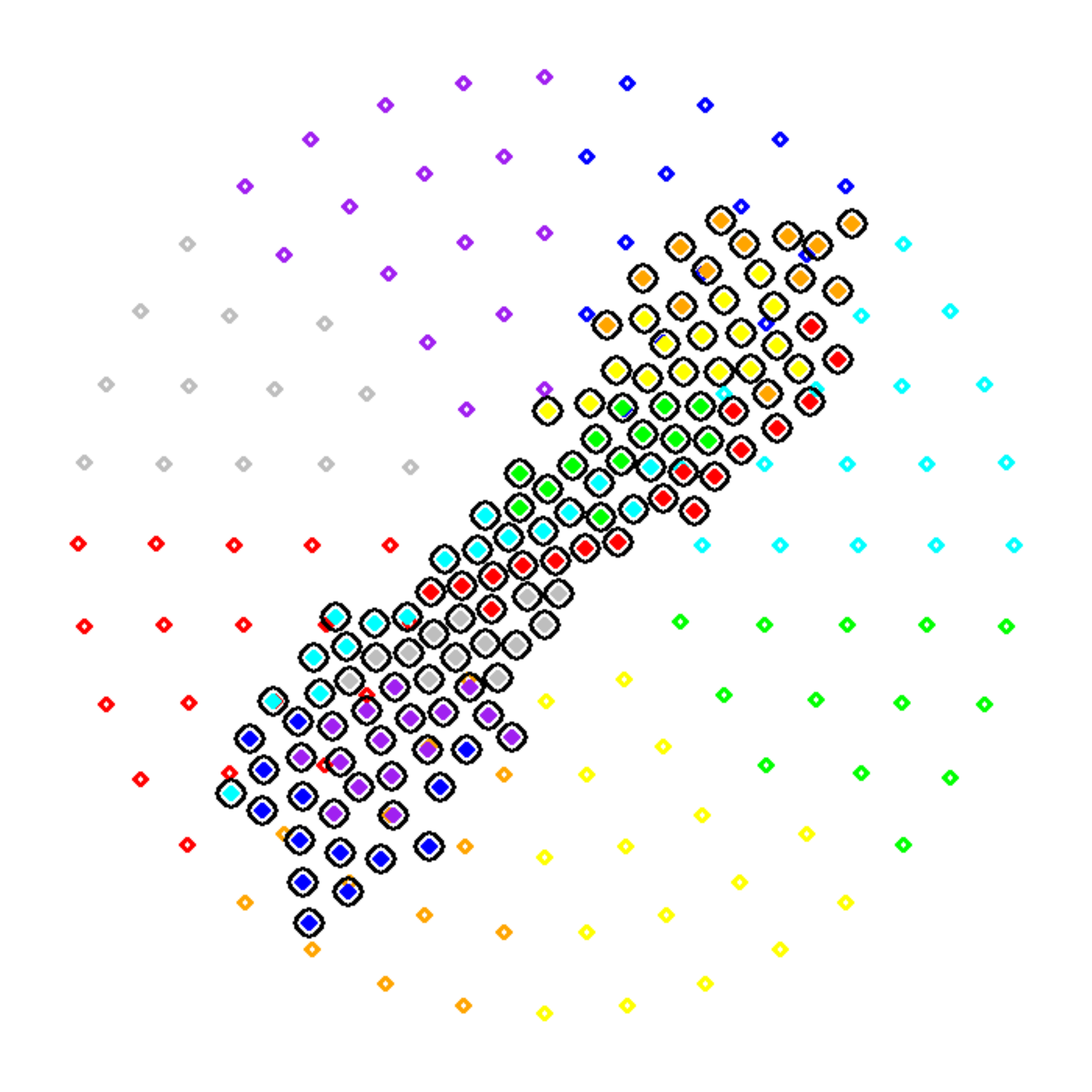}}}
  \caption{Agents resolve the conflicts in the circle scenario,
  where the points of black outline and colored inside are the current positions of the agents,
  and the simply colored points are the target positions of the agents.}
  \label{fig:circle_pass}
\end{figure}

Recall that according to the definition of the safe half-planes,
the local action cell is depressed if there are neighbors approaching.
Therefore, with the same relative position,
it is easier for an agent to ``follow'' a leaving-away neighbor,
if they have the similar preferred trajectories.
As a consequence result, in the case when there are more conflicts,
such as the circle scenario, after gathered around the central area,
instead of squeezing through
(as what happens with the other approaches),
the agents with the LAC based approaches
spin as a whole to resolve the conflicts (Figure~\ref{fig:circle_pass}).

\begin{figure}[!htb]
  \centering
  \includegraphics[width=0.75\textwidth]{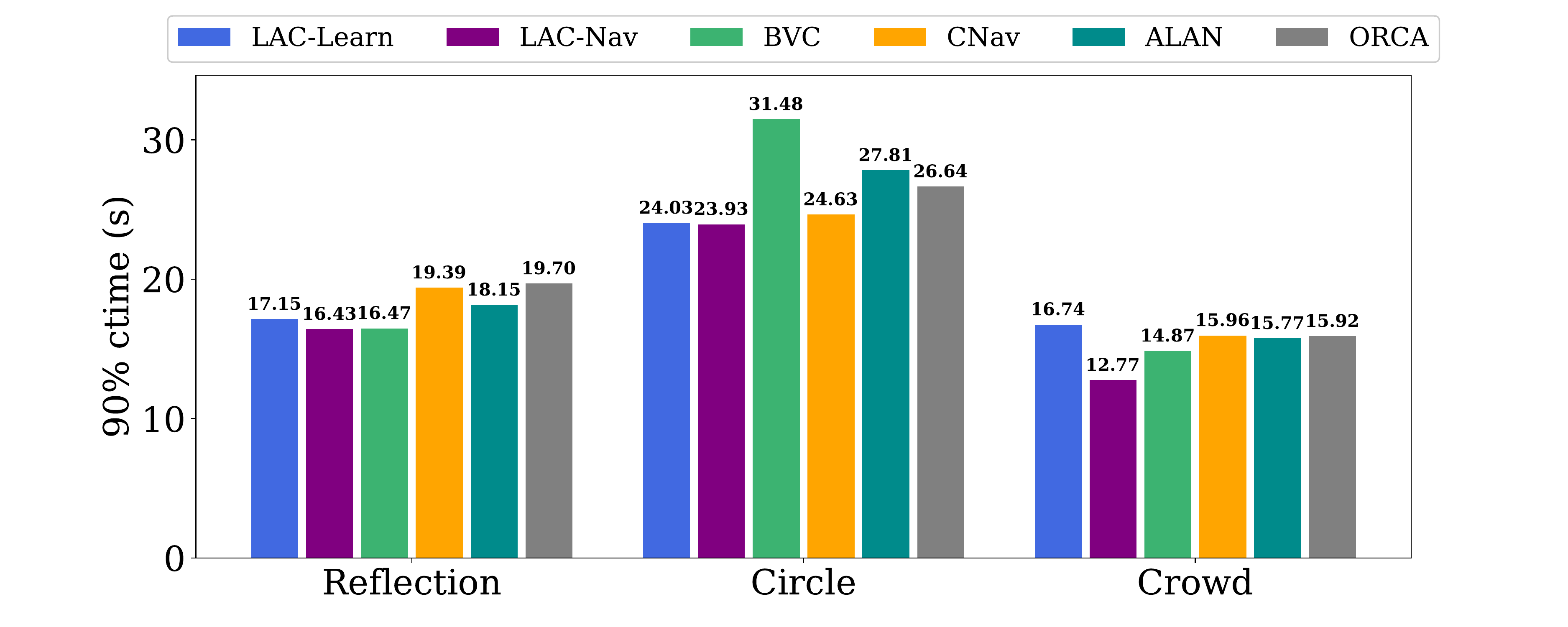}
  \caption{The completion time (in seconds) of the first $90\%$ arrivals.}
  \label{fig:90ctime}
\end{figure}

Although the local action cell can be seen as a variant or extension of the buffered Voronoi cell,
it should be noticed that the LAC based approaches perform distinguishably from BVC, except for
the simple situation such as the reflection scenario.
For the more crowding situations (like the circle scenario and the crowd scenario),
the individual agents with LAC-Nav or LAC-Learn spend more time on average,
while on the other hand the global completion time is shorter.
By investigating into the experiment processes, it can be found that
the LAC based approaches caused less stuck agents than the other approaches did.
This fact can also be revealed by checking
the completion time of the first $90\%$ arrivals (Figure~\ref{fig:90ctime}),
in which the approaches' performances are less distinguishable.

\section{Discussions}

In this work, we introduced the definition of the local action cells, and
proposed two approaches LAC-Nav and LAC-Learn,
of which the efficiency in the completion time have been experimentally demonstrated. 
In order to improve the approaches' performance, besides trying with
different parameter values, there are some natural directions that also extend
the proposed approaches or make a variant.

\paragraph{Adaptive $\lambda$.} Recall that
in the definition of the safe half-plane,
we have set the relax factor as $\lambda := 0.5$.
Intuitively, $\lambda$ indicates how much the agents would like
to compromise in the next move, in order to avoid the collisions that may happen
in a near future. Although it is valid to select any value $[0, 1]$
from the theoretical respect, it should be noted that
a very small $\lambda$ may cause the local action cell being depressed too much,
and a very large $\lambda$ may help little for the long-sighted consideration.
The value $0.5$ is a balanced choice, and it also follows from an important
idea in the reciprocal collision avoidance: each agent take half of the responsibility
to avoid the coming collisions.
However, it will be more interesting if $\lambda$ can be dynamic adjusted
as the agents learned more information about the environment.

\paragraph{Non-uniform $\Delta$.}
In the candidate set $\Delta$ used through this paper,
the angles between any consecutive actions are uniform.
While it is natural to use another uniform candidate set of different size,
say $\Delta$ of size $4$, $12$ or $16$,
it is also valid to include the actions between which
the angles are arbitrary, such as the neighbor-based actions considered in CNav.
In order to prevent the actions being penalized too much,
it should be better to set $\zeta$ close to $1$ or bound maximum penalty,
when the size of $\Delta$ becomes large.

\paragraph{Continuous LAC.} In this paper, we defined
the local action cells as sets of finite number of actions.
However, it may be more natural to consider the continuous
area spanned by the velocities in a cell. There is a direct way
to extend the definition of the local action cell
to include all the linear combinations
between every pair of the adjacent velocities.
Formally speaking, we can define,
\begin{eqnarray*}
\mathcal{C}^*_i &=& \{\lambda_v \cdot v + \lambda_u \cdot u \mid
v~\textrm{and}~u~\textrm{are adjacent in}~\mathcal{C}_i,\\
&&\lambda_v + \lambda_u \leq 1, \lambda_v \geq 0, \lambda_u \geq 0\},
\end{eqnarray*}
which is a continuous area in the velocity space.
With the continuous version the local action cell,
the agents are no longer restricted to select the actions
from $\Delta$, and they can move in any angle as long as
the corresponding velocity has a positive length.

\bibliographystyle{plain}
\bibliography{ref}

\end{document}